\documentclass[aps,showpacs,twocolumn,superscriptaddress,nofootinbib]{revtex4-1}

\usepackage{graphicx}
\usepackage{amssymb}
\usepackage{amsmath}
\begin{document}

\title{Ab-initio Gorkov-Green's function calculations of open-shell nuclei}

\author{V. Som\`a}
\email{vittorio.soma@physik.tu-darmstadt.de}
\affiliation{Institut f\"ur Kernphysik, Technische Universit\"at Darmstadt, 64289 Darmstadt, Germany}
\affiliation{ExtreMe Matter Institute EMMI, GSI Helmholtzzentrum f\"ur Schwerionenforschung GmbH, 64291 Darmstadt, Germany}

\author{C. Barbieri}
\email{C.Barbieri@surrey.ac.uk}
\affiliation{Department of Physics, University of Surrey, Guildford GU2 7XH, UK}

\author{T. Duguet}
\email{thomas.duguet@cea.fr}
\affiliation{CEA-Saclay, IRFU/Service de Physique Nucl\'eaire, F-91191 Gif-sur-Yvette, France}
\affiliation{National Superconducting Cyclotron Laboratory and Department of Physics and Astronomy,
Michigan State University, East Lansing, MI 48824, USA}

\date{\today}

\begin{abstract}
We present results from a new {\em ab-initio} method that uses the self-consistent Gorkov Green's function theory to address truly open-shell systems. The formalism has been recently worked out up to second order and is implemented here in nuclei for the first time on the basis of realistic nuclear forces. We find good convergence of the results with respect to the basis size in $^{44}$Ca and $^{74}$Ni and discuss quantities of experimental interest including ground-state energies, pairing gaps and particle addition/removal spectroscopy.  These results demonstrate that the Gorkov method is a valid alternative to multireference approaches for tackling degenerate or near degenerate quantum systems. In particular, it increases the number of mid-mass nuclei accessible in an ab-initio fashion from a few tens to a few hundreds.
 \end{abstract}


\maketitle

{\it Introduction}. -  The reach of ab-initio nuclear structure calculations has extended tremendously over the last decade. Methods such as coupled-cluster (CC)~\cite{Hagen:2010gd}, in-medium similarity renormalization group (IMSRG)~\cite{Tsukiyama:2010rj} or Dyson self-consistent Green's function~\cite{Barbieri:2009nx} (Dyson-SCGF) have accessed medium-mass nuclei up to $\text{A}\!\sim\!60$ on the basis of realistic two-nucleon (2N) and phenomenological three-nucleon (3N)~\cite{Hagen:2012fb} forces. In their current implementations, such methods are however limited to doubly closed (sub-)shell nuclei and their immediate neighbors~\cite{Dickhoff:2004xx,Jansen:2011cc}. As one increases the nuclear mass, longer chains of truly open-shell nuclei connecting isolated doubly closed-shell ones emerge and cannot be accessed with existing approaches. Many-body techniques that could tackle genuine (at least) singly open-shell systems would immediately extend the reach of ab-initio studies from a few tens to several hundreds of mid-mass nuclei. It is the aim of the present Letter to propose one manageable way to fill this gap.

Typically, open-shell systems can be dealt with via multireference schemes such as, e.g., multireference CC~\cite{MRCC} or configuration interaction techniques based on microscopic one- and two-body components of valence-space interactions~\cite{Tsukiyama:2010rj,Holt:2010yb}. Alternatively, we presently wish to keep the simplicity of a single reference method. This requires however, in any of the approaches mentioned above, to formulate the expansion scheme around a reference state that can tackle Cooper pair instabilities, e.g. to build the correlated state  on top of a Bogoliubov vacuum that already incorporates static pairing correlations. The simplicity of the single reference method can thus be kept at the price of breaking the symmetry associated with particle number conservation. The associated (small) contamination of the results that arises in finite systems eventually calls for the restoration of the broken symmetry~\cite{ring80a}. 
Notice that the use of Bogoliubov-based many-body methods to deal with near-degeneracies and non-dynamical correlations has recently been imported to quantum chemistry and proven to be extremely powerful~\cite{Scuseria:2011pqth}.  Further extension to calculate affinities and ionization energies would require an electron attachment/removal formalism as the one we employ.

The present work discusses the first results obtained by extending Dyson-SCGF theory to the Bogoliubov algebra~\cite{Soma:2011GkvI}, i.e. it carries out the ab-initio Gorkov-SCGF formalism~\cite{Gorkov:1958} in finite nuclei for the first time. A specific benefit of such a method is to provide a way to understand microscopically and quantitatively the processes responsible for the superfluid character of atomic nuclei~\cite{Duguet:2012te}. As a first step, normal and anomalous self-energies are truncated at second-order on the basis of 2N interactions only. This constitutes a Kadanoff-Baym $\Phi$-derivable approximation, i.e. the exact fulfillment of conservation laws is automatically ensured~\cite{Baym:1961zz}. As such, the method involves dressed propagators and is thus intrinsically non perturbative.  In the short-term future, the objectives are to incorporate 3N interactions into the frame and to generalize state-of-the-art Faddeev random-phase approximation (FRPA) truncation scheme~\cite{Barbieri:2000pg,Barbieri:2010eb} to the Gorkov context.

Below, we present converged proof-of-principle calculations of open-shell $^{44}$Ca and $^{74}$Ni nuclei. We employ a next-to-next-to-next-to-leading-order (N$^{3}$LO) 2N chiral interaction~\cite{N3LO:2003}  ($\Lambda_{\chi}=$500\,MeV) complemented by the Coulomb force. The resulting isospin-symmetry breaking interaction is softened using free-space similarity renormalization group (SRG) techniques~\cite{Bogner:2009bt} down to a momentum scale of \hbox{$\lambda = 2.0$ fm$^{-1}$}. After providing basics of Gorkov-SCGF formalism, we illustrate the binding energy convergence with respect to the size of the harmonic oscillator basis used to expand the many-body problem. Then, various observables of experimental interest, i.e. ground-state binding energy, radii and pairing gaps, as well as adjacent isotopes' spectroscopy are discussed. Eventually, the effective neutron shell structure~\cite{baranger70a,Duguet:2011sq,Soma:2011GkvI} is displayed.

{\it Method}. - Results displayed in the present Letter strictly rely on the formalism and the numerical implementation detailed in Refs.~\cite{Soma:2011GkvI} and~\cite{soma12a}, respectively. Given the intrinsic Hamiltonian $H_{\text{int}}\equiv T+V-T_{CM}$, Gorkov-SCGF theory targets the ground state $|  \Psi_0 \rangle$ of the grand-canonical-like potential $\Omega \equiv H_{\text{int}} - \mu \, A$, where $\mu$ is the chemical potential and $A$ the particle-number operator, having the number $\text{A} = \langle  \Psi_0 |  A |  \Psi_0 \rangle$ of particles on average\footnote{Any consideration associated with A=N+Z applies in fact separately to the number of protons Z and to the number of neutrons N.}. The complete one-body information contained in $| \Psi_0 \rangle$ is embodied in a set of four Green's functions\footnote{Vectors and matrices defined on the one-body Hilbert space ${\cal H}_{1}$ are denoted as bold quantities throughout the paper.} $\mathbf{G}^{gg'}(\omega)$ known as Gorkov propagators~\cite{Gorkov:1958}. Their matrix elements read in the Lehmann representation as
\begin{subequations}
\label{eq:leh}
\begin{eqnarray}
\label{eq:leh11}
G^{11}_{ab} (\omega) &=&  \sum_{k} \left\{
\frac{U_{a}^{k} \,U_{b}^{k*}}
{\omega-\omega_{k} + i \eta}
+ \frac{\bar{V}_{a}^{k*} \, {\bar{V}_{b}^{k}}}{\omega+\omega_{k} - i \eta} \right\} \: ,\\
\label{eq:leh12}
G^{12}_{ab} (\omega) &=&   \sum_{k}
\left\{
\frac{U_{a}^{k} \,V_{b}^{k*}}
{\omega-\omega_{k} + i \eta} + \frac{\bar{V}_{a}^{k*} \, {\bar{U}_{b}^{k}}}{\omega+\omega_{k} - i \eta}
\right\}  \, ,\\
\label{eq:leh21}
G^{21}_{ab} (\omega) &=&  \sum_{k}
\left\{
\frac{V_{a}^{k} \,U_{b}^{k*}}
{\omega-\omega_{k}  + i \eta}
+ \frac{\bar{U}_{a}^{k*} \, {\bar{V}_{b}^{k}}}{\omega+\omega_{k} - i \eta}
\right\} \, ,\\
\label{eq:leh22}
G^{22}_{ab} (\omega) &=&   \sum_{k} \left\{
\frac{V_{a}^{k} \,V_{b}^{k*}}
{\omega-\omega_{k} + i \eta}
+  \frac{\bar{U}_{a}^{k*} \, {\bar{U}_{b}^{k}}}{\omega+\omega_{k} - i \eta} \right\} \: .
\end{eqnarray}
\end{subequations}
The poles of the propagators are given by $\omega_{k} \equiv \Omega_k - \Omega_0$, where the index $k$ refers to normalized eigenstates of $\Omega$ that fulfil
\begin{equation}
\label{eq:kapp}
\Omega \, | \Psi_{k} \rangle = \Omega_{k} \, | \Psi_{k} \rangle \: .
\end{equation}
The residue of  $\mathbf{G}^{gg'}(\omega)$ associated with pole $\omega_{k}$ relates to the probability amplitude $\mathbf{U}_k$ ($\mathbf{V}_k$) to reach state $| \Psi_{k} \rangle$ by adding (removing) a nucleon to (from) $| \Psi_{0} \rangle$ on a single-particle state\footnote{The component of vector $\mathbf{U}_k$ associated with a single-particle state $a$ is denoted by $U_{a}^{k}$. Correspondingly, the component associated with the time-reversed state $\bar{a}$  (up to a phase $\eta_a$) is denoted by $\bar{U}_{a}^{k}$~\cite{Soma:2011GkvI}.}. 

Self-consistent, i.e. {\it dressed}, propagators are solutions of Gorkov's equation of motion
\begin{equation}
\label{eq:eigen_uv_bar}
\left.
\left(
\begin{tabular}{c}
\hspace{-0.2cm} $\mathbf{T}  + \mathbf{\Sigma}^{11}(\omega)- \mu \, \mathbf{1} \qquad  \mathbf{\Sigma}^{12}(\omega)$ \\
$\mathbf{\Sigma}^{21}(\omega) \qquad   -\mathbf{T} + \mathbf{\Sigma}^{22}(\omega) + \mu \, \mathbf{1} $
\end{tabular}
\right)
\right|_{\omega_k}
\hspace{-0.2cm}
\left(
  \begin{array}{c}
\hspace{-0.1cm}
 \mathbf{U}  \hspace{-0.1cm}
\\
\hspace{-0.1cm}
\mathbf{V} \hspace{-0.1cm}
  \end{array} \right)_{k}
  \hspace{-0.2cm}
=\omega_{k}
\left(
  \begin{array}{c}
\hspace{-0.1cm} \mathbf{U}\hspace{-0.1cm}  \\
\hspace{-0.1cm}\mathbf{V}\hspace{-0.1cm}
  \end{array} \right)_{k} \: ,
\end{equation}
whose output is the set of $(\mathbf{U}, \mathbf{V})_{k}$ and $\omega_{k}$. Equation~(\ref{eq:eigen_uv_bar}) reads as an eigenvalue problem in which the normal [$\mathbf{\Sigma}^{11}(\omega)$ and $\mathbf{\Sigma}^{22}(\omega)$] and anomalous [$\mathbf{\Sigma}^{12}(\omega)$ and $\mathbf{\Sigma}^{21}(\omega)$] irreducible self-energies act as {\it energy-dependent} potentials. Eventually, the total binding energy of the A-body system is computed via the Koltun-Galitskii sum rule~\cite{Koltun:1972}
\begin{eqnarray}
\label{eq:koltun_gorkov}
E^{\text{A}}_0 &=&
\frac{1}{4 \pi i} \int_{C \uparrow} d \omega \, \text{Tr}_{{\cal H}_{1}}\!\left[ \mathbf{G}^{11} (\omega) \left[ \mathbf{T}
+ \left(\mu + \omega \right) \, \mathbf{1} \right] \right] \, .
\end{eqnarray}
Separation energies between the A-body ground state and eigenstates of $A\pm1$ systems are related to the poles $\omega_k$ through 
\begin{eqnarray}
E_k^{\pm} \equiv  \mu \pm \omega_k 
&=&   \pm \left[\langle \Psi_k | H_{\text{int}} |  \Psi_k \rangle - \langle  \Psi_0 | H_{\text{int}} |  \Psi_0 \rangle\right] \nonumber \\
&& \mp \mu \left[\langle  \Psi_k | A |  \Psi_k \rangle - (\text{A}\pm 1)\right] \,\,\,, \label{eq:epmk1}  
\end{eqnarray}
where the error associated with the difference between the average number of particles in state $|  \Psi_k \rangle$ and the targeted particle number $\text{A}\pm1$ is taken care of by the last term of Eq.~\eqref{eq:epmk1}. Spectroscopic factors associated with the {\it direct} addition and removal of a nucleon are defined as
\begin{eqnarray}
SF_{k}^{+} &\equiv& \text{Tr}_{{\cal H}_{1}}\!\left[\mathbf{U}_{k} \mathbf{U}^{\dagger}_{k}\right]  \; \; \text{and} \; \;
SF_{k}^{-} \equiv \text{Tr}_{{\cal H}_{1}}\!\left[\mathbf{V}^{\ast}_{k}\mathbf{V}^{T}_{k}\right]  .  \label{SF_Gorkov2}
\label{eq:SF}
\end{eqnarray}
In open-shell nuclei, the odd-even staggering of nuclear masses is a fingerprint of pairing correlations and offers, through finite odd-even mass difference formulae, the possibility to extract the pairing gap. The most commonly used~\cite{duguet02b} three-point-mass difference formula $\Delta^{(3)}_{n}(\text{A})$ equates the pairing gap with the Fermi gap in the one-nucleon addition/removal spectra $E_k^{\pm}$, e.g. $\Delta^{(3)}_{n}(\text{A}) \equiv (-1)^{\text{A}} [E_0^{+}-E_0^{-}]/2$. One-body observable such as mass or charge radii can be easily computed from $\mathbf{G}^{11}(\omega)$~\cite{Soma:2011GkvI}. Moreover, effective single-particle energies (ESPE) introduced by Baranger as centroids $e^{\text{cent}}_a$ of one-nucleon addition and removal spectra $E_k^{\pm}$ can be naturally computed in the present context~\cite{Soma:2011GkvI}. Last but not least, the normal self-energy $\mathbf{\Sigma}^{11}(\omega)$ is identified with the microscopic nucleon-nucleus optical potential~\cite{Capuzzi96,Waldecker:2011by}, allowing for the computation of scattering states~\cite{Barbieri:2005NAscatt}.

Proceeding to an actual Gorkov-SCGF calculation requires to truncate the diagrammatic expansion of the four self-energies ${\bf \Sigma}^{gg'}(\omega)$. As opposed to perturbation theory, the expansion involves skeleton diagrams expressed in terms of dressed propagators solution of Eq.~\eqref{eq:eigen_uv_bar}. Such a key feature of {\it self-consistent} Green's function methods allows the re-summation of self-energy insertions to all orders and makes the method intrinsically non-perturbative and iterative. In the present application, self-consistency is limited to the static, i.e. energy-independent, part~\cite{Polls94} ${\bf \Sigma}^{gg'}(\infty)$ of the self energy. This constitutes the so-called "sc0" approximation that grasps the dominant fraction of self-consistency effects at a tractable numerical cost~\cite{Barbieri:2010eb,soma12a}. At first order in vacuum interactions, Eq.~(\ref{eq:eigen_uv_bar}) reduces to an ab-initio Hartree-Fock-Bogoliubov (HFB) problem with energy-independent normal and anomalous self-energies accounting for Hartree-Fock and Bogoliubov diagrams, respectively. In the present application, both first- and second-order diagrams are included~\cite{Soma:2011GkvI}.

Gorkov's equation~(\ref{eq:eigen_uv_bar}) can be transformed into a energy-independent eigenvalue problem of larger dimensionality~\cite{Soma:2011GkvI} and solved iteratively. The algorithm includes a readjustment of the chemical potential to ensure that proton and neutron numbers are correct in average. The most dramatic aspect of the implementation consists of handling the increased dimensionality, i.e. the growing number of poles in ${\bf G}^{gg'}(\omega)$ [Eq.~(\ref{eq:leh})] with iterations, by means of a Lanczos algorithm. All numerical aspects will be reported on in a forthcoming publication~\cite{soma12a}.

\begin{figure}[t]
\includegraphics[width=0.85\columnwidth,clip=true]{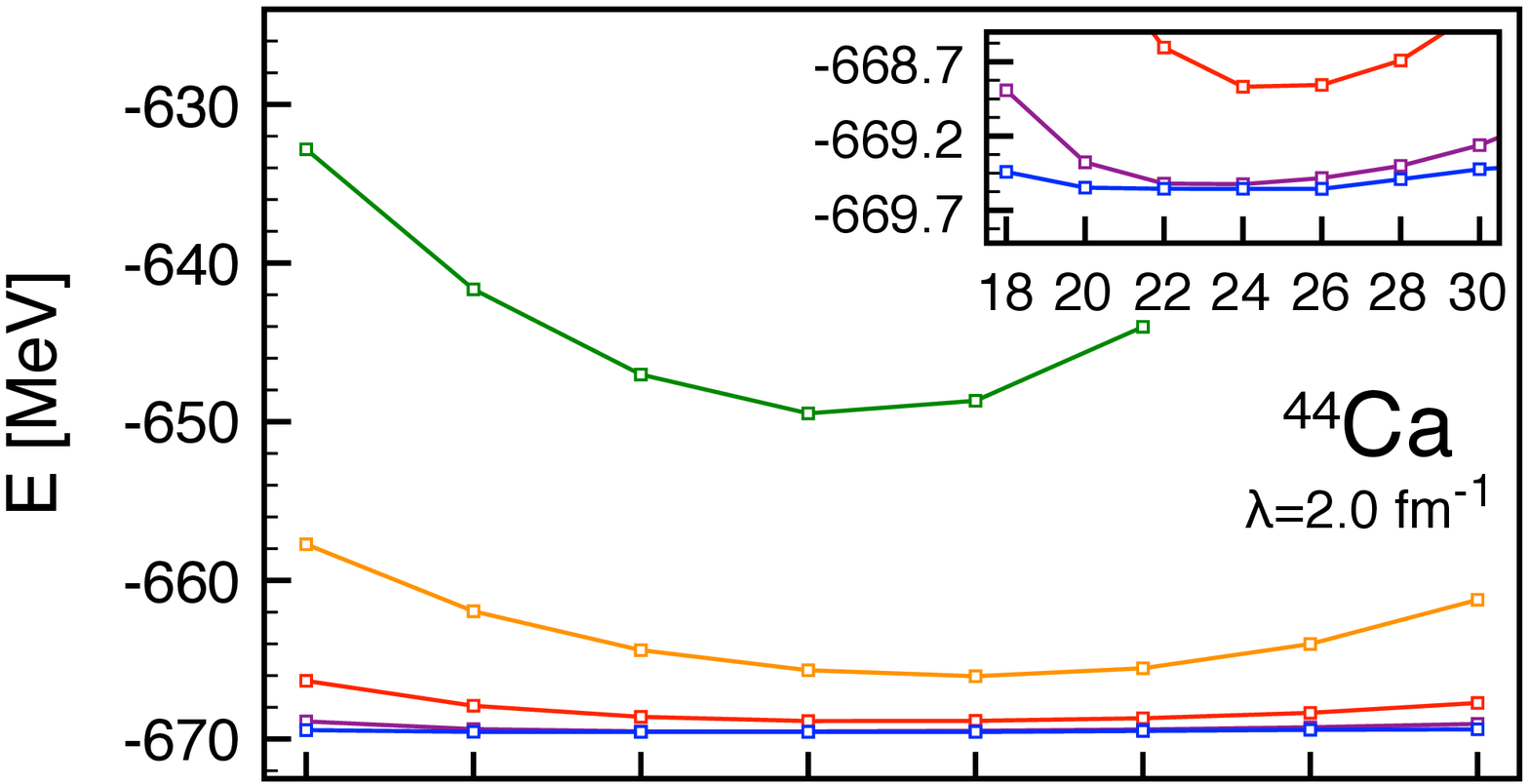}
\vskip 0.1cm
\includegraphics[width=0.85\columnwidth,clip=true]{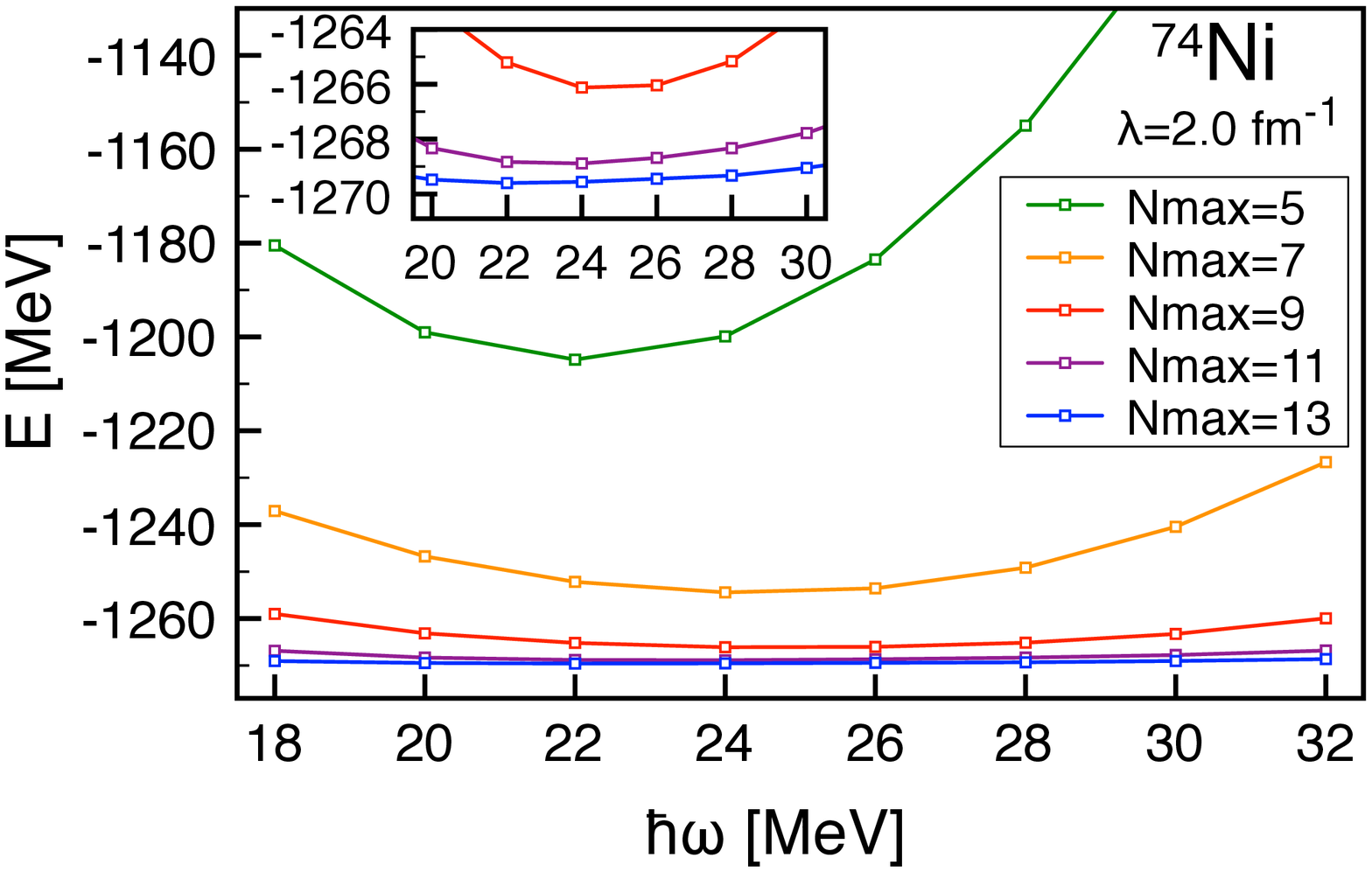}
\caption{(Color online)  Binding energy for $^{44}$Ca (upper panel) and $^{74}$Ni (lower panel) as a function of the harmonic oscillator spacing $\hbar\omega$ and for an increasing size $N_{max} \equiv$ max $(2n+l)$ of the single-particle model space. Results are from (sc0) second-order Gorkov-SCGF calculations. The inserts show a zoom on the most converged results. 
}
\label{fig:convergence}
\vspace{-0.5cm}
\end{figure}

{\it Results}. - Figure~\ref{fig:convergence} displays the binding energy of $^{44}$Ca and $^{74}$Ni as a function of the harmonic oscillator spacing $\hbar\omega$ and for an increasing size $N_{max} \equiv$ max $(2n+l)$ of the single-particle model space. The convergence pattern obtained here on the basis of a soft 2N interaction is similar to those generated for doubly-closed shell nuclei with currently available ab-initio methods. Overall, convergence is well attained for $N_{max}=13$. In $^{44}$Ca, going from $N_{max}=11$ to $N_{max}=13$ lowers the minima by just a few keV. Also, the binding energy calculated for  $N_{max}=13$ varies by less than 200\,keV over a wide range of $\hbar\omega$ values. In $^{74}$Ni, going from $N_{max}=11$ to $N_{max}=13$ yields an additional $600$\,keV, while scanning a large range of oscillator frequencies only changes the binding energy by about $1$\,MeV.

Table~\ref{tab:numbers} lists the results obtained for various observables of interest in the ground state of $^{44}$Ca and $^{74}$Ni. The values quoted are extrapolated to infinite oscillator basis size using the method proposed in Ref.~\cite{furnstahl12a}. At this point, results are mostly illustrative because of the lack of 3N forces. The latter play a key role in the saturation of nuclear matter such that omitting it generates too much binding and too small nuclei when using soft 2N interactions~\cite{Bogner:2009bt}. The neglect of 3N forces also induces too small pairing gaps as a result of the too low density of states in the nucleon addition and removal spectra (see below). It is our short-term objective to add 3N forces to the present theoretical scheme.

\begin{figure}[h]
\includegraphics[width=0.85\columnwidth,clip=true]{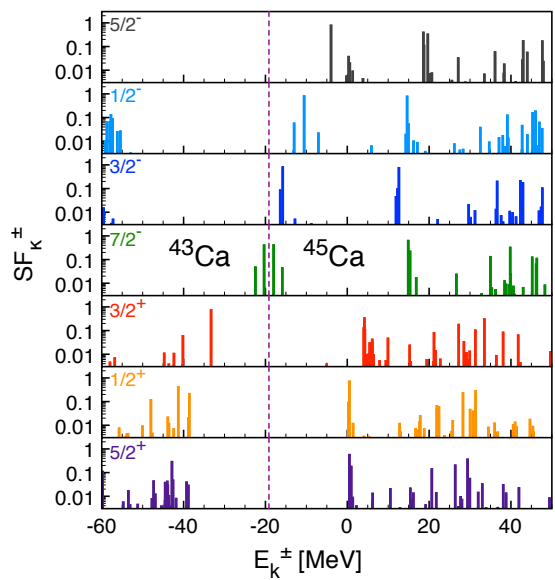}
\caption{(Color online) One-neutron addition and removal spectral strength distributions in $^{44}$Ca obtained from second-order (sc0) Gorkov-SCGF calculations.
 For each final state in $^{43}$Ca (left to the dashed line) and in $^{45}$Ca (right to the dashed line), the spectroscopic factor is plotted as a function of its separation energy to the ground state of $^{44}$Ca. Energies above 0~MeV correspond to $n$+$^{44}$Ca scattering states~\cite{Barbieri:2005NAscatt}. Final states with different $J^{\pi}$ values are separated for clarity. Results correspond to the minimum of the convergence plots shown in Fig.~\ref{fig:convergence}. Although center of mass motion is subtracted by using $H_{\text{int}}$, the {\it variation} of that correction going from A to A$\pm$1 is neglected. The associated error is small in such medium-mass nuclei~\cite{soma12a}.
}
\label{fig:spectral_functionCa44}
\vspace{-0.2cm}
\end{figure}

Figure~\ref{fig:spectral_functionCa44} displays one-neutron addition and removal spectral strength distributions (SSD)  in $^{44}$Ca. Results are shown over a large range of final states in $^{43}$Ca and $^{45}$Ca characterized by spectroscopic factors as small as $2.10^{-3}$ (i.e. 0.2$\%$). One observes a fragmentation of the spectroscopic strength that is characteristic of correlated many-body systems. Overall the pattern is similar to the one found in doubly-magic nuclei~\cite{Barbieri:2009nx}. Close to the Fermi energy, however, one notices a feature that is unique to open-shell nuclei, i.e. the $7/2^{-}$ strength is equally fragmented into additional and removal channels, which results in the fact that both $^{43}$Ca and $^{45}$Ca ground-states have angular momentum and parity $J^{\pi}=7/2^{-}$. Such a fragmentation reflects static pairing correlations that manifest themselves as a result of emerging degeneracies in the ground state of open-shell nuclei. It is the main strength of Gorkov-SCGF theory to explicitly handle such degeneracies and resulting pairing correlations.

\begin{figure}[h]
\includegraphics[width=0.85\columnwidth,clip=true]{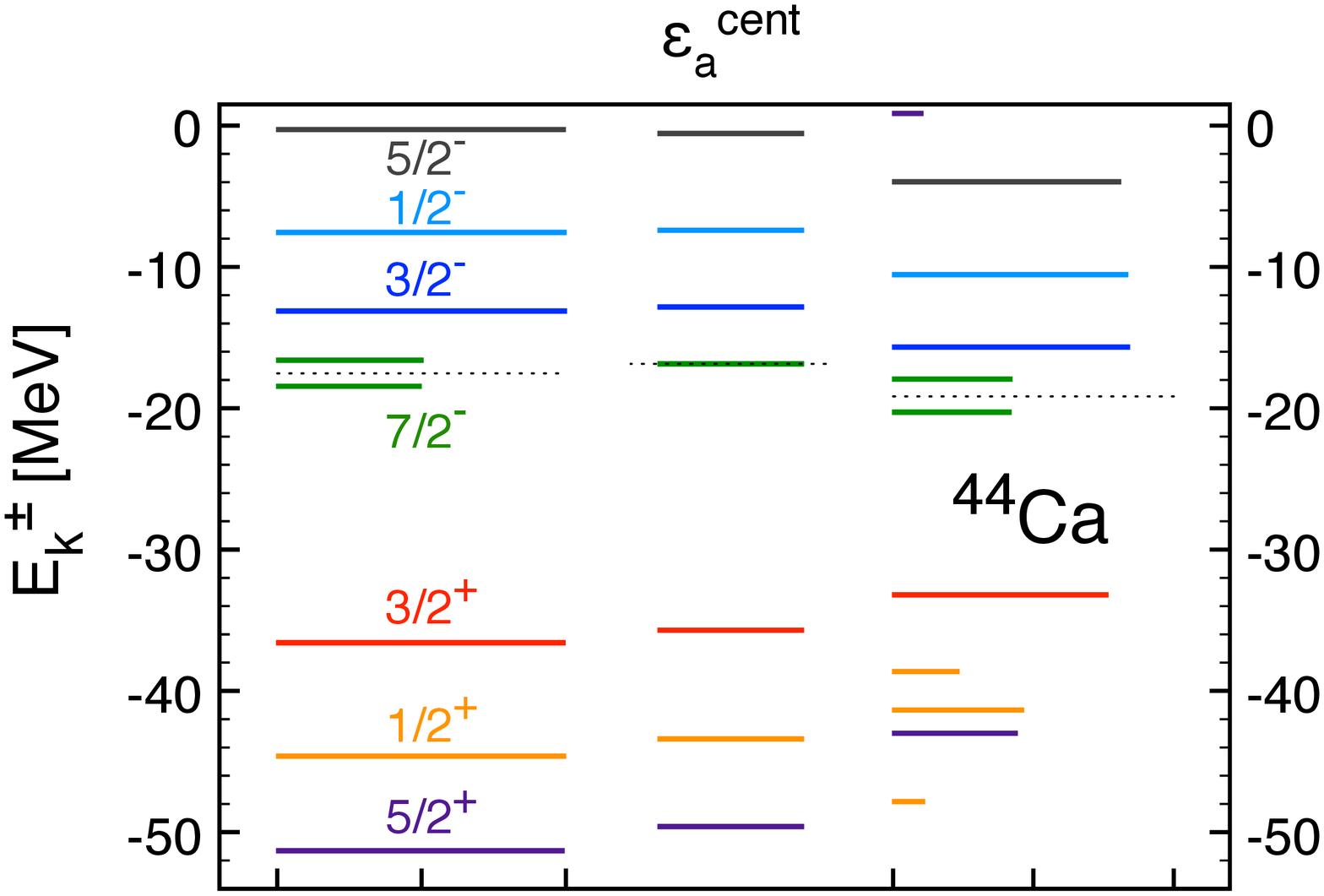}
\vskip 0.2cm
\includegraphics[width=0.85\columnwidth,clip=true]{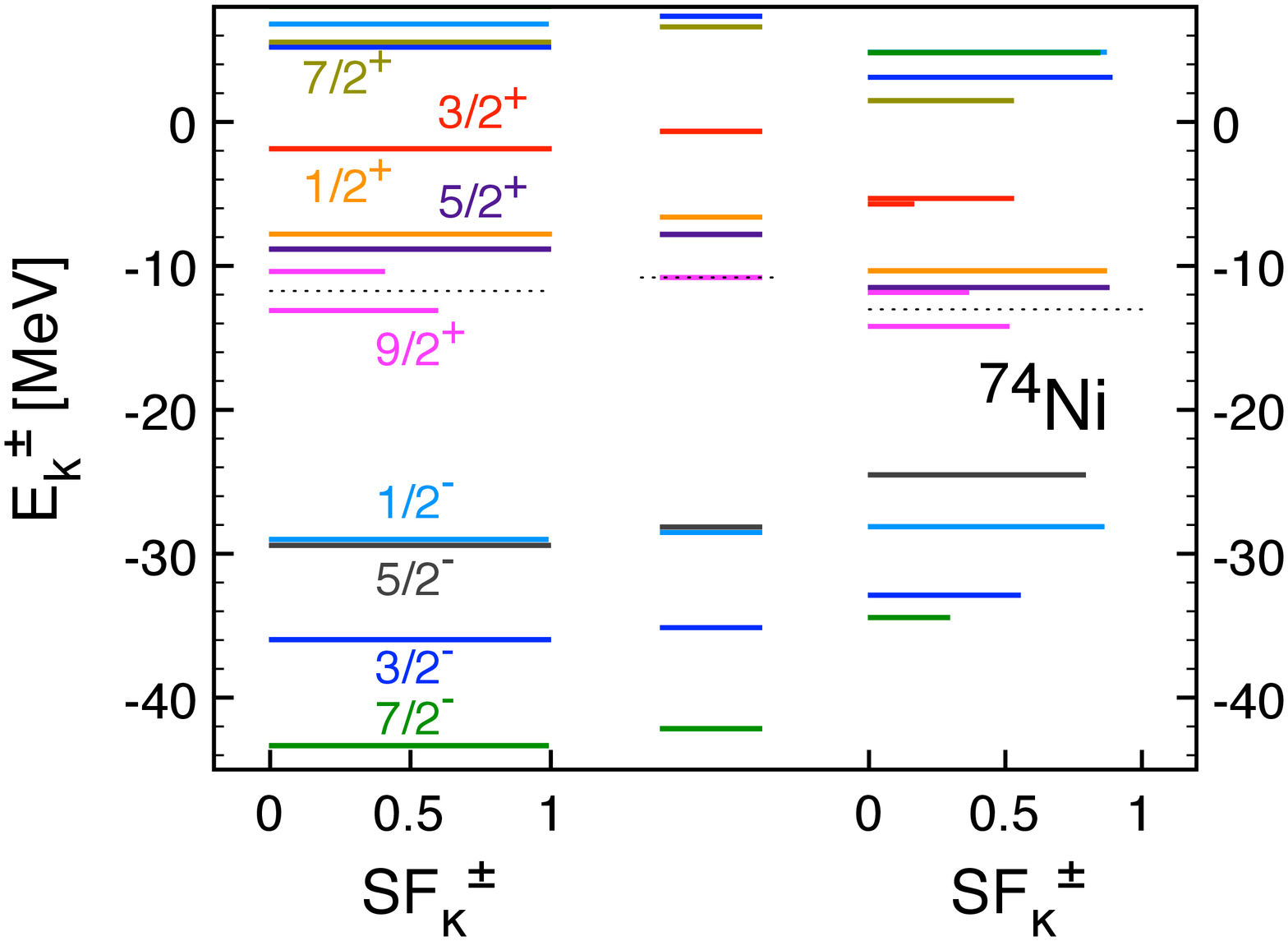}
\caption{(Color online) Left: One-neutron addition and removal spectral strength distribution obtained from first-order (HFB) Gorkov-SCGF calculations. Right: same as left panel for second-order (sc0) calculations. Center: Baranger ESPEs reconstructed from second-order (sc0) Gorkov-SCGF calculations. Upper panel: $^{44}$Ca. Lower panel: $^{74}$Ni.
}
\label{fig:Ca44_Ni74SPEs}
\end{figure}

The right column in the upper panel of Fig.~\ref{fig:Ca44_Ni74SPEs} supplies a zoom of Fig.~\ref{fig:spectral_functionCa44} around the Fermi energy for states with spectroscopic factors larger than $10^{-1}$ (i.e. $10\%$). The left column provides the same quantities for first-order (i.e. HFB) calculations. Last but not least, the center column displays effective single-neutron energies. The same information is provided for $^{74}$Ni in the lower panel of Fig.~\ref{fig:Ca44_Ni74SPEs}.

The main fragmentation of the strength is absent from first-order calculations, i.e. it is due to dynamical correlations that come in at second order and that are qualitatively the same as for closed-shell nuclei. Contrarily, the fragmentation of the strength in the vicinity of the Fermi energy into two peaks of (essentially) equal strength is qualitatively accounted for at first order and thus relates predominantly to static pairing correlations. Quantitatively, the energy spacing between the two low-lying $7/2^-$ states in the SSD of $^{44}$Ca is increased by second-order effects. This is in contrast to the behavior of $^{74}$Ni where the separation of the low-lying $9/2^+$ states is instead decreased. Given that such a spacing is equal to (twice) the pairing gap, one concludes that the coupling of  Cooper pairs to non-collective fluctuations may already affect pairing correlations in open-shell nuclei in either way. A detailed study of such a feature is left to a forthcoming publication~\footnote{A quantitative treatment of nuclear superfluidity through ab-initio approaches requires to treat the coupling of the Cooper pair to {\it collective} density, spin and isospin fluctuations~\cite{gori05,Duguet:2012te}. In the present context, this necessitates the implementation of the (generalized) FRPA expansion scheme~\cite{Barbieri:2000pg,Barbieri:2010eb}.}. Back to the overall spectrum, one observes that the position of the dominant peak of a given $J^{\pi}$ value is significantly modified by second-order effects such that the corresponding spectrum is more compressed than at first order. However, it is still significantly too spread out compared to experiment due to missing 3N forces and the lack of coupling to collective fluctuations.

Effective single-particle energies recollect the fragmented strength~\cite{baranger70a,Duguet:2011sq,Soma:2011GkvI} from both one-nucleon addition and removal channels. Many-body correlations are largely screened out from ESPEs, which picture the averaged single-nucleon dynamics inside the correlated system. Two different features are identifiable in the ESPE spectrum $e^{\text{cent}}_{a}$ when compared to observable one-nucleon addition and removal spectra $E^{\pm}_k$. The ESPE $e^{\text{cent}}_{1f_{7/2}}$ ($e^{\text{cent}}_{1g_{9/2}}$) located at the Fermi energy recollects the strength of the two equally important $7/2^{-}$ ($9/2^+$) states. Other ESPEs recollect the strength of a low-lying dominant peak and of a highly fragmented strength distributed at higher excitation energies such that they move away from the Fermi energy to closely match first-order, i.e. HFB, peaks. This is consistent with the fact that ESPEs inform on the averaged, mean-field-like, one nucleon dynamics.

\begin{table}[t]              
\begin{ruledtabular}
\begin{tabular}{lcccc}
                  &~ $E^{\text{A}}_{0}$ & $\Delta^{(3)}_{n}(\text{A})$ & $r_{rms}$ \\
\hline                 

$^{44}$Ca      &     $-669.6(1)$   &  $1.16 $    &  $2.48$     & \\
$^{74}$Ni      &     $-1269.7(2)$   &  $1.17(1)$    &  $2.75$     & \\ 
\end{tabular}
\end{ruledtabular}
 \caption[]{Binding energy (MeV), neutron pairing gap (MeV) and matter root mean square radius (fm).
Results are from second-order (sc0) Gorkov-SCGF calculations 
and are extrapolated to infinite oscillator basis size using the method of Ref.~\cite{furnstahl12a}. The extrapolation error is indicated only when it is
bigger than the last digit shown.
    }
\label{tab:numbers}
\vspace{-0.2cm}
\end{table}


{\it Conclusions}. - We have presented the results of the first-ever {\em ab-initio} calculations of medium-mass (truly) open-shell nuclei. Such calculations are based on the implementation of self-consistent Gorkov Green's function theory on the basis of realistic nuclear interactions. Taking $^{44}$Ca and $^{74}$Ni as test cases, we have demonstrated the good convergence of the results with respect to the basis size and discussed several quantities of experimental interests including ground-state energies, pairing gaps and particle addition/removal spectroscopy. Such calculations increase the reach of ab-initio calculations in the mid-mass region tremendously and are now being performed systematically over long isotopic and isotonic chains~\cite{soma12a}. The short-term objectives are to incorporate three-nucleon interactions 
into the framework and to extend state-of-the-art Faddeev-random-phase-approximation truncation scheme from doubly-closed shell nuclei to open-shell nuclei, i.e to the present Gorkov context.

{\it Acknowledgements}. - This work was supported by the United Kingdom Science and Technology Facilities Council (STFC) under Grants ST/I003363 and ST/J00005, by the DFG through grant SFB 634 and by the Helmholtz Alliance Program, contract HA216/EMMI. V. S. acknowledges support from Espace de Structure Nucl\'eaire Th\'eorique (ESNT) at CEA/Saclay.
Calculations were performed using HPC resources at CCRT under GENCI Grant 50707.


\bibliography{bisodu}

\end{document}